\documentclass[fleqn]{2020SCGE}
\usepackage{bm}

\usepackage{multirow}

\begin{document}

\ensubject{}%¶þŒ¶Ñ§¿Æ
%\luntan

\ArticleType{Letter to the Editor}%{Short Communications}%%ÀžÄ¿% Letter to the Editor   %Erratum    %Highlight   %Comments  %Invited Review   %News \& Views
\SpecialTopic{}%{Special Topic: Physics Behind the $H_0$ Tension}%×šÌâÃû³Æ
\Year{2020}
\Month{September}
\Vol{63}
\No{9}
\DOI{10.1007/s11433-019-1514-0}
\ArtNo{290406}
\ReceiveDate{December 8, 2019}
\AcceptDate{January 20, 2020}
\OnlineDate{May 8, 2020}

\title{Inflation model selection revisited after a 1.91\% measurement \\of the Hubble constant}{Inflation model selection revisited after a 1.91\% measurement of the Hubble constant}

\author[1]{Rui-Yun Guo}{}
\author[2]{Jing-Fei Zhang}{}
\author[2]{Xin Zhang}{zhangxin@mail.neu.edu.cn} %\Authorfootnote 

\AuthorMark{R.Y. Guo}%\authorcr????????

\AuthorCitation{R.-Y. Guo, J.-F. Zhang, and X. Zhang}

\address[1]{Department of Physics, College of Sciences, Xi'an Technological University, Xi'an 710021, China}
\address[2]{Department of Physics, College of Sciences, Northeastern University, Shenyang 110819, China}

\maketitle

%%%%%%%%%%%%%%%%%%%%%%%%%%%%%%%%%%%%%%%%%%%%%%%%%%%%%%%
%%% The main text. ÕýÎÄ²¿·Ö
%±ížñºÍÍŒ±í¹«ÊœµÄÒýÓÃÒªÓÃ\cref{fig1}

%\tableofcontents  %Ë÷ÒýÄ¿ÂŒ
\vspace*{-6mm}

\begin{multicols}{2}

%\section{Introduction}

\noindent The recent local measurement of the Hubble constant based on the distance-ladder method has reached a 1.91\% precision \cite{Riess:2019cxk}, but this result is in tension with the early-universe measurements at the more than 4$\sigma$ level, bringing a crisis to the contemporary cosmology. In addition to the end-to-end test of the $\Lambda$CDM model in the late universe, it is also of great interest to see how the local $H_0$ measurement affects the determination of the primordial power spectra, and further to test the influences for the inflation model selection. Here, we constrain the primordial power spectra of scalar and tensor perturbations by using a series of observational data, including the Planck 2015 cosmic microwave background (CMB) temperature and polarization power spectra data \cite{Aghanim:2015xee}, the Planck 2018 lensing power spectrum data \cite{Aghanim:2019ame}, the BICEP2/Keck Array CMB B-mode data \cite{Ade:2018gkx}, and also the prior of optical depth $\tau=0.054\pm0.007$ \cite{Aghanim:2018eyx}, as well as the late-universe measurements (baryon acoustic oscillations and type Ia supernovae). In particular, we use the latest 1.91\% measurement of the Hubble constant, $H_{0}=74.03\pm1.42$ km s$^{-1}$ Mpc$^{-1}$ \cite{Riess:2019cxk}, in this cosmological test. We find that considering the latest local $H_{0}$ measurement in the data combination will lead to a larger fit value of $n_{\rm s}$. With the addition of the latest local measurement of $H_{0}$, it is found that the natural inflation model is totally excluded at the $2\sigma$ level, the Starobinsky $R^{2}$ inflation model is marginally favored at around the $2\sigma$ level, and the spontaneously broken SUSY inflation model is the most favored model.

We consider the $\Lambda$CDM+$r$+$N_{\rm eff}$ model, where $r$ denotes the tensor-to-scalar ratio at the pivot scale $k_{*} = 0.002$ Mpc$^{-1}$ and $N_{\rm eff}$ is the effective number of relativistic degrees of freedom. 
%In addition to $r$ and $N_{\rm eff}$, other parameters of this model include the baryon density ($\Omega_{\rm b}h^{2}$), the cold dark matter density ($\Omega_{\rm c}h^{2}$), the ratio between the sound horizon size and the angular diameter distance at the last-scattering epoch ($\theta_{\ast}$), the optical depth to reionization ($\tau$), the scalar spectral index ($n_{\rm s}$), and the amplitude of the primordial scalar power spectrum ($A_{\rm s}$). The priors of all these parameters are the same as those in ref.~\cite{Ade:2015xua}. 
%The number of e-folds $N$ is uniformly taken as $N\in [50, 60]$. 
%To derive the posterior probability distributions of parameters, we use the Markov-chain Monte Carlo (MCMC) sampler {\tt CosmoMC}~\cite{Lewis:2002ah}.
Three data combinations, i.e., Planck+BK15+BAO, Planck+BK15+BAO+SN, and Planck+BK15+BAO+SN+$H_{0}$, are considered. 
Here, ``Planck'' denotes the Planck CMB data including the temperature, polarization, and lensing data plus the $\tau$ prior, ``BK15'' denotes the new BICEP2/Keck observations up to and including the 2015 observing season, and BAO and SN denote the latest baryon acoustic oscillation and type Ia supernova observations, respectively.
We use the three data combinations to constrain cosmological parameters.

%The detailed constraint results are shown in Table~\ref{table1}. 
%For the scalar spectral index and tensor-to-scalar ratio, we obtain $n_{\rm s}\!=\!0.9642\pm0.0069$ and \linebreak $r_{0.002}\!\!<\!\!0.057$ from Planck+BK15+BAO; we obtain $n_{\rm s}$ \linebreak $=0.9651\pm0.0069$ and $r_{0.002}\!<\!0.057$ from Planck +BK15+BAO+SN; and we obtain $n_{\rm s}=0.9794^{+0.0056}_{-0.0057}$ and $r_{0.002}<0.061$ from Planck+BK15+BAO+SN+$H_0$. Obviously, the SN data almost do not affect the fit results of $n_\textmd{s}$ and $r$, because the two parameters are purely from the early-universe physics. It is found that the Harrison-Zel'dovich (HZ) scale-invariant spectrum with $n_{\rm s} = 1$ is excluded at the $5.06\sigma$ level using the Planck+ BK15+BAO+SN data. But when the local measurement of $H_0$ is added into the data combination, we find that the value of $n_{\rm s}$ is increased from $n_{\rm s}=0.9651\pm0.0069$ to $n_{\rm s}=0.9794^{+0.0056}_{-0.0057}$, and thus the HZ scale-invariant spectrum is excluded at the $3.68\sigma$ level. Meanwhile, the upper limit of $r_{0.002}$ is increased from $r_{0.002}<0.057$ to $r_{0.002}<0.061$. We can see that, when the local measurement of $H_0$ is considered in the cosmological fit, the constraint on $n_\textmd{s}$ is evidently changed, but the constraint on $r_{0.002}$ is only slightly changed.

\begin{figure}[H]
\centering\vspace*{1mm}
\includegraphics{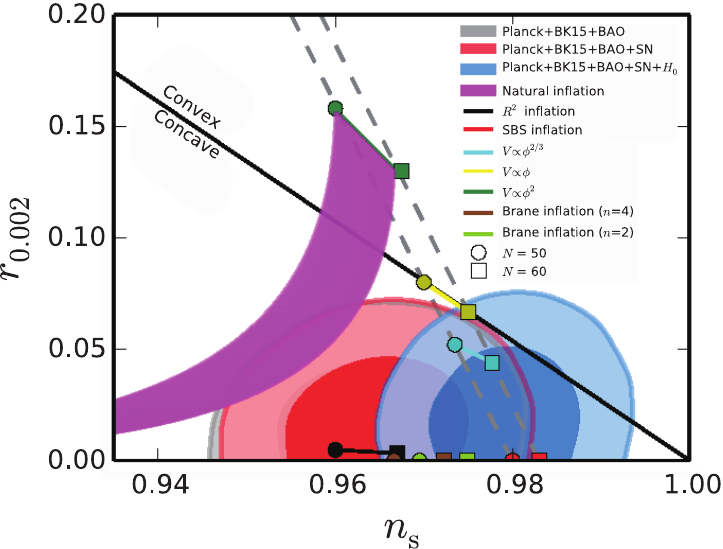}\vspace*{-2mm}
\caption{(Color online) Two-dimensional contours ($1\sigma$ and $2\sigma$) for $n_{\rm s}$ and $r_{\rm 0.002}$ using the Planck+BK15+BAO, Planck+BK15+BAO+SN, and Planck+BK15+BAO+SN+$H_{0}$ data, compared to the theoretical predictions of selected inflation models.}
\label{fig1}
\end{figure}

We show the constraint results in Figure~\ref{fig1}. We can clearly see that the contour of $(n_{\rm s}, r_{0.002})$ from the Planck \linebreak  +BK15+BAO+SN data almost totally overlaps with that from the Planck+BK15+BAO data. Thus, in the following, we only discuss the constraint results from the Planck+BK15+BAO+SN data and the Planck+BK15+BAO +SN+$H_{0}$ data, to show the impacts of the local measurement of $H_{0}$ on inflation model selection. Comparing the red and blue contours, we find that, once the local measurement of $H_{0}$ is added to the Planck+BK15+BAO+SN data, an obvious right shift of $n_{\rm s}$ is yielded, resulting in some great changes for the inflation model selection.

Now we compare the theoretical predictions of several typical inflation models with the fit results of $(n_{\rm s}, r_{0.002})$. First, we discuss the case of Planck+BK15+BAO+SN, i.e., the red contours. We can clearly see that in this case the inflation models with concave potentials are favored by the current observations at the 2$\sigma$ level. The natural inflation model is only marginally favored, because only a small part of it still resides in the 2$\sigma$ edge of this case. For the chaotic inflation models with a monomial potential, we find that the $\phi^2$ model is excluded at much more than 2$\sigma$ level, the $\phi$ model is excluded at around 2$\sigma$ level, and the $\phi^{2/3}$ model is only excluded at the 1$\sigma$ level and it still resides in the 2$\sigma$ region. The SBS inflation model is also only favored at around the 2$\sigma$ level because it is on the edge of the 2$\sigma$ region.
The Starobinsky $R^2$ inflation model and the brane inflation model (with $n=2$ and $n=4$) are well consistent with the Planck+BK15+BAO+SN data. In this case, the most favored model is the Starobinsky $R^2$ model.

Next, we discuss the comparison of predictions of inflation models with the blue contours corresponding to the case of considering the addition of the local measurement of $H_0$ in the data combination. We can see that, in this case, neither the concave potential nor the convex potential is excluded by the observational data. It seems that, however, when making a comparison for the two types of potentials, the inflation models with a concave potential is more favored by the observations. We find that now the natural inflation model is excluded at more than 2$\sigma$ level. For the chaotic inflation models, the $\phi^2$ model is excluded at much more than 2$\sigma$ level, the $\phi$ model is marginally favored at the 2$\sigma$ level (the case with $N=60$ still resides on the verge of 2$\sigma$ region), and the $\phi^{2/3}$ model is marginally favored at around 1$\sigma$ level. Dramatic change happens to the situation of the Starobinsky $R^2$ inflation model, because now it is only marginally favored by the observations at around the 2$\sigma$ level (the case with $N=60$ still resides on the verge of 2$\sigma$ region). The brane inflation model (with $n=2$ and $n=4$) is still consistent with the observational data at the 2$\sigma$ level. In this case, the SBS inflation model becomes the most favored model by the observations.

%\section{Conclusion}

In summary, we constrain the $\Lambda$CDM+$r$+$N_{\rm eff}$ model by using the current observations including the latest local measurement of the Hubble constant. We discuss the issue of inflation model selection based on the constraint results. The main conclusion of this work is still accordant with the previous studies \cite{Tram:2016rcw,Guo:2017qjt,Guo:2018uic}, i.e., the consideration of the latest local measurement of $H_0$ leads to a great change for the inflation model selection. In the case of considering the $H_0$ measurement, the Starobinsky $R^2$ inflation model is only marginally favored at around the 2$\sigma$ level, and the most favored model is the SBS inflation model.

\Acknowledgements{This work was supported by the National Natural Science Foundation of China (Grant Nos.~11975072, 11875102, 11835009, and 11690021), the Liaoning Revitalization Talents Program (Grant No. XLYC1905011), and the Fundamental Research Funds for the Central Universities (Grant No. N2005030).}

\end{multicols}
\end{document}